\documentclass[sigconf]{acmart}

\usepackage{booktabs} 
\usepackage{hyperref}

\setcopyright{none}
\setcopyright{rightsretained}

\acmDOI{}
\def\acmDOI#1{\def\@acmDOI{#1}}
\acmISBN{}

\acmConference[LBRS@RecSys'18, October 2-7, 2018, Vancouver, BC, Canada]{In Proceedings of the Late-Breaking Results 
track part of the Twelfth ACM Conference on Recommender Systems 
(LBRS@RecSys'18), Vancouver, BC, Canada, October 6, 2018, 2 pages.}
\copyrightyear{2018}

\begin{document}
\title{Can we leverage rating patterns from traditional users to enhance recommendations for children?}

\author{Ion Madrazo Azpiazu, Michael Green, Oghenemaro Anuyah, Maria Soledad Pera}
\affiliation{%
  \institution{People and Information Research Team -- Department of Computer Science}
  \streetaddress{Boise State University}
  \city{Boise}
  \state{Idaho}
  \postcode{83702}
}
\email{{ionmadrazo, michaelgreen1, oghenemaroanuyah, solepera}@boisestate.edu}

\renewcommand{\shortauthors}{Madrazo Azpiazu et al.}

\begin{abstract}
Recommender algorithms performance is often associated with the availability of sufficient historical rating data. 
Unfortunately, when it comes to 
children, this data is seldom available. 
In this paper, we report on 
an initial analysis conducted to examine the degree to which data about traditional users, i.e., adults, 
can be leveraged to enhance the recommendation process for children.
\end{abstract}

%
%

\keywords{Children; analysis; transfer learning; recommendations}

\maketitle

\section{Introduction}
The success of recommendation systems which capture user behavioral patterns to offer relevant recommendations is often correlated with the availability of sufficient historical data in the form of ratings.
When dealing with traditional users (e.g., adults), obtaining this information is usually less problematic than the case of non-traditional users (e.g., children). Data to inform design and evaluation of algorithms targeting traditional audiences 
is often publicly available (e.g., MovieLens). However, when it comes to children it is difficult to obtain sufficient historical rating data due to 
privacy rules like COPPA or GDPR. 

The idea of transfer learning has commonly been applied to crossdomain recommendations \cite{cantador2015cross}. In this paper, we instead argue in favor of integrating information about one target audience to inform and enhance the recommendation process for another: in the absence of sufficient information for children, we infer knowledge from ratings of traditional users to improve recommendations.
To do so, we explore the MovieLens \cite{guo2013novel} dataset and one created from children's ratings made available through Dogo Movies \cite{dogomovies}. 
We use a number of research question to drive the empirical analysis conducted across different recommender algorithms and datasets to see if it is possible to enhance rating predictions for children's movies by leveraging rating patterns from traditional users. 


\section{Experimental Set Up}
We detail our experimental framework below.
\footnote{Scripts to generate datasets and reproduce experiments can be found in \url{https://doi.org/10.18122/cs_scripts/6/boisestate}.} 

\textbf{Data}. Due to the lack of datasets comprised of items rated by children we created  \textbf{Dogo}, which uses data from Dogo Movies~\cite{dogomovies}, a site where children can rate and review children's movies. 
To inform our analysis, we also 
use the well-known MovieLens dataset (\textbf{ML1M}) \cite{guo2013novel}. See Table \ref{tab:results} for statistics on these datasets.

\textbf{Algorithms}. We consider the following algorithms: 
\begin{itemize}
\item  \textit{Item-Item} (\textbf{II}), a popular item-based collaborative filter~\cite{sarwar2001item}; based on cosine similarity. 
\item  \textit{User-User} (\textbf{UU}), a traditional user-based collaborative filter~\cite{herlocker2002empirical}; user similarity based on Pearson coefficient. 

\item \textit{BiasedMF} (\textbf{MF}), 
a well-known strategy based on matrix facorization \cite{koren2009matrix}; using 100 training iterations, regularization of .06, and a learning rate of .07. 
\end{itemize}

\textbf{Metric}. To quantify recommender performance, we use RMSE. 
\section{Discussion \& Analysis}

\begin{figure}[!ht]
\begin{center}
\resizebox{0.5\textwidth}{!}{
\includegraphics[scale = 0.5]{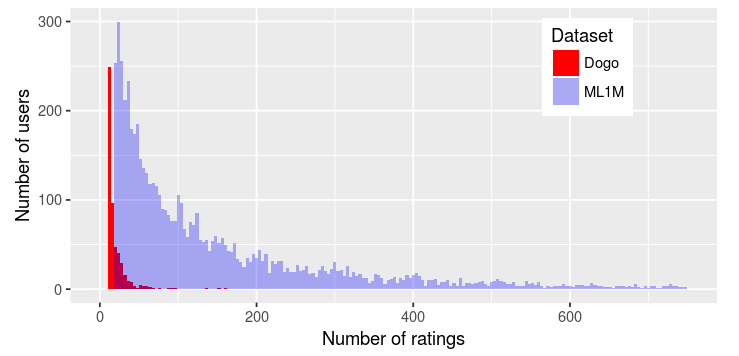}
}
\end{center}
\caption{User-rating activity in Dogo and ML1M.  
}
 \label{fig:dogo_dist}
\end{figure}

\textbf{Are baselines applicable to offer recommendations to children?} To answer this, we applied II, UU, and MF to Dogo, for contextualizing algorithm performance. As shown in Table \ref{tab:results}, ML1M and Dogo yield similar RMSE scores. Given the differences in rating distribution among users in Dogo and ML1M (see Figure \ref{fig:dogo_dist}), we argue that 
algorithm performance on Dogo warrants further examination. Thus, we created subsets of Dogo where we varied the minimum number of items a user must rate in order to be considered in the corresponding subset. We do this until we reach a minimum of 20 ratings per user, following the premise of ML1M. 

We see that regardless of the algorithm, RMSE scores increase across Dogo datasets. We attribute this to the decrease in users that fulfill minimum requirements. This translates into less instances an algorithm can use to create neighborhoods, 
which in turn affects overall performance. Also, as the minimum number of ratings per user increases, the number of users that can be served by the algorithm is reduced by approximately 97\%.

\begin{table*}[!ht]
\caption{Performance analysis based on RMSE. X in Dogo\_x is the minimum number of items a user rates to be in the dataset. [.] is the neighborhood size that yielded best performance for UU and II; for MF, it is the number of latent factors. \textit{Tr} refers to training set, \textit{Te} to test set, and \textit{K+} to ratings provided by ML1M users who rated both children and non-children's movies.} 
\label{tab:results}
\resizebox{\textwidth}{!}{
\begin{tabular}{|l|c|c|c|c|c|c|c|}
\hline
\textbf{Dataset}                      & \textbf{Users} & \textbf{Items} & \textbf{Ratings}  & \textbf{Min \# of ratings} & \textbf{UU}     & \textbf{II}     & \textbf{MF}     \\ \hline \hline
ML1M                                  & 6,040          & 3,706          & 1,000,209         & 20                         & 0.905 {[}80{]}  & 0.876 {[}80{]}  & 0.852 {[}120{]} \\ \hline
Dogo\_2                               & 5,496          & 2,054          & 28,368            & 2                          & 0.861 {[}150{]} & 0.899 {[}250{]} & 0.823 {[}60{]}  \\ \hline
Dogo\_10                              & 613            & 1,589          & 11,271            & 10                         & 0.884 {[}50{]}  & 0.914 {[}50{]}  & 0.829 {[}120{]} \\ \hline
Dogo\_20                              & 156            & 1,223          & 5,392             & 20                         & 0.894 {[}150{]} & 0.924 {[}50{]}  & 0.836 {[}120{]} \\ \hline \hline
ML1M \& Dogo\_2\_Tr::Dogo\_2\_Te       & 11,126::3,946  & 5,255::1,182   & 1,017,249::10,057 & 20::2                      & 1.309 {[}50{]}  & 1.031 {[}50{]}  & 0.880 {[}120{]} \\ \hline
ML1M \& Dogo\_10\_Tr::Dogo\_10\_Te     & 6,656::613     & 4,854::925     & 1,007,014::4,302  & 20::10                     & 1.278 {[}50{]}  & 0.981 {[}50{]}  & 0.874 {[}120{]} \\ \hline
ML1M \& Dogo\_20\_Tr::Dogo\_20\_Te     & 6,196::156     & 4,572::683     & 1,003,463::2,032  & 20::20                     & 1.252 {[}50{]}  & 1.012 {[}50{]}  & 0.899 {[}120{]} \\ \hline \hline
ML1M\_K+ \& Dogo\_2\_Tr::Dogo\_2\_Te   & 11,028::3,946  & 1,782::1,182   & 177,855::10,057   & 20 \& 2::2                    & 1.231 {[}50{]}  & 1.011 {[}50{]}  & 0.873 {[}120{]} \\ \hline
ML1M\_K+ \& Dogo\_10\_Tr::Dogo\_10\_Te & 6,558::613     & 1,381::925     & 167,620::4,302    & 20 \& 10::10                   & 1.235 {[}50{]}  & 0.973 {[}50{]}  & 0.870 {[}120{]} \\ \hline
ML1M\_K+ \& Dogo\_20\_Tr::Dogo\_20\_Te & 6,098::156     & 1,099::683     & 164,069::2,032    & 20 \& 20::20                   & 1.224 {[}50{]}  & 1.012 {[}50{]}  & 0.892 {[}120{]} \\ \hline
\end{tabular}
}
\end{table*}

\textbf{Can we use rating patterns from adults to inform recommendations for children?} We simulated the recommendation process using variations of Dogo that incorporated ML1M, i.e.,  we combined ML1M with 60\% of Dogo, and used the remaining 40\% for testing purposes. We exclusively use Dogo for testing to quantify effect on recommender algorithm performance for children when introducing rating instances that have not been generated by them. 

As shown in Table \ref{tab:results}, we find that irrespective of the minimum number of ratings in Dogo and the recommender algorithm used, RMSE scores obtained as a result of integrating rating patterns from traditional users are higher than the one obtained by using their corresponding Dogo counterparts. We hypothesize that this is because the ratio of users who rated adult movies in the ML1M dataset were more than those that rated children's movies (8:1).

While we expected algorithm performance to be improved by the availability of more data, failing to capture rating patterns that mimic those of children introduced noise. This led us to question if performance could be improved by focusing on traditional users that have rated children's movies. 

\textbf{Can we find a special group of traditional users that can aid recommendations for children?} To identify a special group of traditional users, we selected users in ML1M that have rated at least 2 children's movies. We infer that these users are similar to Dogo users, being that they have taken interest in movies for children. Given all ratings provided by these users (i.e., both for children and non-children's movies), we combined them with 60\% of variations of Dogo (according to the minimum ratings). Again, we rely on instances based on just Dogo for testing, as we want to investigate the performance of recommender algorithms when integrating ratings based on users that perform similar to children. 


As showcased in Table \ref{tab:results}, When compared to results from the previous analysis, in most cases, the performance of recommender algorithms improves (t-test, $p$ $<$ 0.05). We credit this improvement to the fact that, in this case, recommender algorithms are able to capture more relevant rating patterns among this special group of users, as opposed to other users in ML1M. We do observe, however, that RMSE scores yielded in this experiment are higher than those obtained by the Dogo counterparts. As a result of further analyzing these datasets, we see that ratings in Dogo are consistently high (i.e., 4 to 5), whereas rating generated by  traditional users in ML1M are distributed along the entire rating spectrum (i.e., 1 to 5). We attribute the low recommender algorithm performance (i.e., approximately 6\% decrease) to this difference in rating patterns.  

\section{Conclusions and Future Work}
We have presented the latest results conducted to validate the use of historical data from traditional users to inform and ultimately enhance the recommendation task for non-traditional users. Insights from our analysis reveal that even though we leverage ML1M to enhance Dogo recommendations, we infer that improved performance is not obtained due to a difference in behavioral patterns between traditional and non-traditional users. 
Based on results from our experiments, we will continue our quest to identify special groups of traditional users, as well as incorporate meta data such as reviews and time-based information, to further enhance recommendations for non-traditional users. 

\begin{acks}
 Work partially funded by NSF Award 1565937.

\end{acks}

\bibliographystyle{ACM-Reference-Format}
\bibliography{poster-bibliography}


\begin{thebibliography}{6}


\ifx \showCODEN    \undefined \def \showCODEN     #1{\unskip}     \fi
\ifx \showDOI      \undefined \def \showDOI       #1{#1}\fi
\ifx \showISBNx    \undefined \def \showISBNx     #1{\unskip}     \fi
\ifx \showISBNxiii \undefined \def \showISBNxiii  #1{\unskip}     \fi
\ifx \showISSN     \undefined \def \showISSN      #1{\unskip}     \fi
\ifx \showLCCN     \undefined \def \showLCCN      #1{\unskip}     \fi
\ifx \shownote     \undefined \def \shownote      #1{#1}          \fi
\ifx \showarticletitle \undefined \def \showarticletitle #1{#1}   \fi
\ifx \showURL      \undefined \def \showURL       {\relax}        \fi
\providecommand\bibfield[2]{#2}
\providecommand\bibinfo[2]{#2}
\providecommand\natexlab[1]{#1}
\providecommand\showeprint[2][]{arXiv:#2}

\bibitem[\protect\citeauthoryear{Cantador, Fern{\'a}ndez-Tob{\'\i}as,
  Berkovsky, and Cremonesi}{Cantador et~al\mbox{.}}{2015}]%
        {cantador2015cross}
\bibfield{author}{\bibinfo{person}{Iv{\'a}n Cantador}, \bibinfo{person}{Ignacio
  Fern{\'a}ndez-Tob{\'\i}as}, \bibinfo{person}{Shlomo Berkovsky}, {and}
  \bibinfo{person}{Paolo Cremonesi}.} \bibinfo{year}{2015}\natexlab{}.
\newblock \showarticletitle{Cross-domain recommender systems}.
\newblock In \bibinfo{booktitle}{\emph{Recommender Systems Handbook}}.
  \bibinfo{publisher}{Springer}, \bibinfo{pages}{919--959}.
\newblock


\bibitem[\protect\citeauthoryear{Dogo}{Dogo}{2018}]%
        {dogomovies}
\bibfield{author}{\bibinfo{person}{Dogo}.} \bibinfo{year}{2018}\natexlab{}.
\newblock \bibinfo{title}{{Dogo Movies: Movie reviews by kids for kids}}.
\newblock \bibinfo{howpublished}{\url{https://www.dogomovies.com/ }. Accessed:
  July 2018}.
\newblock


\bibitem[\protect\citeauthoryear{Guo, Zhang, and Yorke-Smith}{Guo
  et~al\mbox{.}}{2013}]%
        {guo2013novel}
\bibfield{author}{\bibinfo{person}{G. Guo}, \bibinfo{person}{J. Zhang}, {and}
  \bibinfo{person}{N. Yorke-Smith}.} \bibinfo{year}{2013}\natexlab{}.
\newblock \showarticletitle{A Novel Bayesian Similarity Measure for Recommender
  Systems}. In \bibinfo{booktitle}{\emph{Proceedings of the 23rd International
  Joint Conference on Artificial Intelligence (IJCAI)}}.
  \bibinfo{pages}{2619--2625}.
\newblock


\bibitem[\protect\citeauthoryear{Herlocker, Konstan, and Riedl}{Herlocker
  et~al\mbox{.}}{2002}]%
        {herlocker2002empirical}
\bibfield{author}{\bibinfo{person}{Jon Herlocker}, \bibinfo{person}{Joseph~A
  Konstan}, {and} \bibinfo{person}{John Riedl}.}
  \bibinfo{year}{2002}\natexlab{}.
\newblock \showarticletitle{An empirical analysis of design choices in
  neighborhood-based collaborative filtering algorithms}.
\newblock \bibinfo{journal}{\emph{Information retrieval}} \bibinfo{volume}{5},
  \bibinfo{number}{4} (\bibinfo{year}{2002}), \bibinfo{pages}{287--310}.
\newblock


\bibitem[\protect\citeauthoryear{Koren, Bell, and Volinsky}{Koren
  et~al\mbox{.}}{2009}]%
        {koren2009matrix}
\bibfield{author}{\bibinfo{person}{Yehuda Koren}, \bibinfo{person}{Robert
  Bell}, {and} \bibinfo{person}{Chris Volinsky}.}
  \bibinfo{year}{2009}\natexlab{}.
\newblock \showarticletitle{Matrix factorization techniques for recommender
  systems}.
\newblock \bibinfo{journal}{\emph{Computer}} \bibinfo{number}{8}
  (\bibinfo{year}{2009}), \bibinfo{pages}{30--37}.
\newblock


\bibitem[\protect\citeauthoryear{Sarwar, Karypis, Konstan, and Riedl}{Sarwar
  et~al\mbox{.}}{2001}]%
        {sarwar2001item}
\bibfield{author}{\bibinfo{person}{Badrul Sarwar}, \bibinfo{person}{George
  Karypis}, \bibinfo{person}{Joseph Konstan}, {and} \bibinfo{person}{John
  Riedl}.} \bibinfo{year}{2001}\natexlab{}.
\newblock \showarticletitle{Item-based collaborative filtering recommendation
  algorithms}. In \bibinfo{booktitle}{\emph{10th International Conference on
  World Wide Web (WWW)}}. ACM, \bibinfo{pages}{285--295}.
\newblock


\end{thebibliography}

\end{document}